\documentstyle[preprint,aps,epsfig]{revtex}

\newcommand{\beq}{\begin{eqnarray}}
\newcommand{\eeq}{\end{eqnarray}}
\newcommand{\beqs}{\begin{eqnarray}}
\newcommand{\eeqs}{\end{eqnarray}}
\newcommand{\bary}{\begin{array}}
\newcommand{\eary}{\end{array}}

\newcommand{\h}{h}

\newcommand{\BR}{\mbox{BR}}

\newcommand{\figpos}{p}      
\newcommand{\spaAR}{6.5}     
\newcommand{\heightAR}{14.0} 
\newcommand{\widthAR}{14.0}  

\newcommand{\andthis}{~~~~~\mbox{and}~~~~~}
\newcommand{\orthis}{~~~~~\mbox{or}~~~~~}
\newcommand{\withthis}{~~~~~\mbox{with}~~~~~}

\def\eq#1{{\rm eq.}~(\ref{#1})}
\def\eqs#1{{\rm eqs.}~(\ref{#1})}
\def\Eq#1{{\rm Eq.}~(\ref{#1})}
\def\apks{{a_{\psi K_S}}}
\def\eps{\varepsilon}
\def\epsK{\eps_K}
\def\Re{{\cal R}e}

\def\npb#1{Nucl.\ Phys.\ {\bf B\,#1}}

\def\plb#1{Phys.\ Lett.\ {\bf B\,#1}}

\def\prd#1{Phys.\ Rev.\ {\bf D\,#1}}
\def\prl#1{Phys.\ Rev.\ Lett. {\bf#1}}

\def\epjc#1{Eur.~Phys.~J.\ {\bf C\,#1}}

\def\progtp#1{Prog.\ Th.\ Phys.\ {\bf #1}}

\def\hepa#1{{\tt hep-ph/#1}}
\def\hepb#1{{[{\tt hep-ph/#1}]}}
\def\hepea#1{{\tt hep-ex/#1}}

\def\gsim{\ \rlap{\raise 3pt \hbox{$>$}}{\lower 3pt \hbox{$\sim$}}\ }
\def\lsim{\ \rlap{\raise 3pt \hbox{$<$}}{\lower 3pt \hbox{$\sim$}}\ }

\def\putFig#1#2#3#4#5#6#7 
{
 \begin{figure}[\figpos]
 \LARGE
 $#7$
 \begin{center}
 \mbox{\epsfig{figure=#1.eps,angle=0,width=#3cm,height=#4cm}}
 $#6$
 \end{center}
 \vspace{1.5cm}
 \caption{#2}
 \label{#1}
 \end{figure}
}

\begin{document}

\preprint{\vbox{\hbox{WIS-01/5/MAR-DPP}
                \hbox{hep-ph/0103299}}}

\title{~\\ Constraining Models of New Physics \\
        in Light of Recent Experimental Results on $\apks$} 

\author{Sven Bergmann and Gilad Perez \\ 
        \small \it Department of Particle Physics,
        Weizmann Institute of Science,
        Rehovot 76100, Israel}
\maketitle

\begin{abstract}
  We study extensions of the Standard Model where the charged current
  weak interactions are governed by the CKM matrix and where all
  tree-level decays are dominated by their Standard Model
  contribution. We constrain both analytically and numerically the
  ratio and the phase difference between the New Physics and the
  Standard Model contributions to the mixing amplitude of the neutral
  $B$ system using the experimental results on $R_u$, $\Delta
  m_{d,s}$, $\eps_K$ and $\apks$. We present new results concerning
  models with minimal flavor violation and update the relevant
  parameter space.  We also study the left-right symmetric model with
  spontaneously broken CP, probing the viability of this model in view
  of the recent results for $\apks$ and other observables.
\end{abstract}

\vspace{1cm}

\section{Introduction}

For more than 30 years the only unambiguous indication for CP
violation (CPV) was found in the neutral kaon system. Even though the
CP violating parameter
\beq \label{epsK}
\epsK = (2.280 \pm 0.013) \times 10^{-3} e^{i \phi_\eps}
~~~\mbox{with}~~~
\phi_\eps \simeq \pi/4 
\eeq
has been measured rather accurately~\cite{PDG}, hadronic uncertainties
obscure the determination of the fundamental CKM parameters
\cite{CKM}.  However, there is growing evidence that CPV also occurs
in the $B$ meson system.  Specifically, indications for a CP asymmetry
in $B_d \to J/\psi \, K_S$ decays have been reported by the
ALEPH~\cite{ALEPH}, BaBar~\cite{BaBarexp}, BELLE~\cite{Belle},
CDF~\cite{CDF} and OPAL~\cite{OPAL} collaborations.  Combining all the
measurement gives:
\beq \label{apsiKs}
a_{\psi K_S} = 0.51 \pm 0.18 \,. 
\eeq

It is essential to understand the impact of the present and upcoming
experimental data on the SM and its extensions. In this paper we focus
on a large class of NP models where the charged current weak
interactions are governed by a $3 \times 3$ unitary matrix, just like
in the CKM picture.  Moreover, we make the reasonable assumption that
all tree-level decays are dominated by their SM contribution. However,
flavor changing neutral current processes are sensitive to New Physics
(NP) because, in the SM, they appear only at the loop level.

After a brief update on the SM picture, we study in detail, both
analytically and numerically, the allowed parameter space of the ratio
and the phase difference between the NP and the SM contributions to
the mixing amplitude of the neutral $B$ system.  We use the
experimental results on $R_u$, $\Delta m_{d,s}$, $\eps_K$ and $\apks$
for the generic NP models. Besides the two general assumptions
mentioned above we introduce additional constraints that apply to
specific NP models.  First, we assume that the ratio between the mass
differences of the $B_d$ and the $B_s$ neural meson system is
unaffected by NP, which is valid, {\it e.g.}, in models with Minimal
Flavor Violation (MFV). Second, we impose a small CP violating phase,
as for example in left-right symmetric (LRS) models with spontaneous
CPV.

We devote special attention to models of MFV. We show that the three
parameters $\rho$, $\eta$ and $F_{tt}$ can be obtained analytically
from the experimental results on $R_u$, $\Delta m_d$ and $\eps_K$.
Additional constraints on the parameter space arise from the
measurement of $\apks$ and the upper bound on $\Delta m_d / \Delta
m_s$.

Finally, we investigate the impact of the uncertainty of various
parameters in LRS models with spontaneous CPV on recent claims that
these models would be ruled out by a significant CP asymmetry in $B_d
\to J/\psi \, K_S$ decays.

\section{Standard Model picture}
\label{SM}

CPV appears naturally in the Standard Model with three generations of
quarks. It can be attributed to a single phase, $\delta_{KM}$,
appearing in the CKM matrix that describes the weak charged-current
interactions.  Among the four Wolfenstein parameters
$(\lambda,A,\rho,\eta)$ that provide a convenient description of the
CKM matrix~\cite{WOLF} only two have been determined with a good
accuracy~\cite{PDG}, namely
\beq \label{lambdaA}
\lambda = 0.2237 \pm 0.0033 \andthis A = 0.819 \pm 0.040 \,.
\eeq
The other two parameters $(\rho,\eta)$ describing the apex of the {\it
  unitarity triangle\/}, that follows from the unitarity relation
%
$
V_{ud} V_{ub}^* + V_{cd} V_{cb}^* + V_{td} V_{tb}^* = 0 \,,
$
%
have a rather large uncertainty. As a consequence the CP violating
phase $\delta_{KM} = \arctan(\eta/\rho)$ of the SM is relatively
weakly constrained.

In the following we review briefly how the allowed region for the two
Wolfenstein parameters $\rho$ and $\eta$ is obtained (for more details
see {\it e.g.}~\cite{BaBar,Buras2001} and references therein).  At
present these parameters are experimentally constrained by the
measurement of $|V_{ub}/V_{cb}|$, the observed $B_d^0 - \bar B_d^0$
mixing parameter $\Delta m_d$, the lower bound on the $B_s^0 - \bar
B_s^0$ mixing parameter $\Delta m_s$, the measurement of $\epsK$ and
the combined result for $a_{\psi K_S}$.  Note that for the purpose of
this paper it is sufficient to work to leading order in $\lambda^2$
and we therefore neglect all subdominant contributions.

The distance from the origin to the apex $(\rho,\eta)$ of the rescaled
unitarity triangle,
\beq \label{Rb}
\sqrt{\rho^2 + \eta^2}  \simeq
R_u \equiv \left|{V_{ub}^* V_{ud} \over V_{cb}^* V_{cd}}\right| \,,
\eeq
is proportional to $|V_{ub}/V_{cb}|$, which dominates its uncertainty.
The ratio $|V_{ub}/V_{cb}|$ can be determined from various tree-level
$B$ decays (see {\it e.g.}~\cite{CKM2000} and Tab 1.\,), restricting
$R_u$ to the interval
\beq \label{Rb-interval}
R_u \in [0.34,~0.43] \,.
\eeq

The distance between the apex of the rescaled unitarity triangle and
the point $(1,0)$ in the $\rho-\eta$ plane is given by:
\beq \label{Rt}
\sqrt{(\rho-1)^2 + \eta^2} \simeq
R_t \equiv \left|{V_{tb}^* V_{td} \over V_{cb}^* V_{cd}}\right| \,.
\eeq
$R_t$ can be determined from the experimental results for mass
differences $\Delta m_q$ in the $B_q^0 - \bar B_q^0$ ($q=d, s$)
systems by means of the SM predictions~(see {\it
  e.g.}~\cite{BaBar,Buras2001}):
\beq \label{SM-DelMb}
\Delta m_q = {G_F^2 \over 6 \pi^2} \eta_B m_{B_q} m_W^2 f^2_{B_q}
 B_{B_q} S_0(x_t) \cdot |V_{tq} V_{tb}^*|^2 \,.
\eeq
Here $G_F$ is the Fermi constant, $\eta_B$ is a QCD correction factor,
$m_{B_q}$ is the $B_q^0$ mass, $m_W$ is the $W$-boson mass, $f_{B_q}$
is the $B_q$ decay constant, $B_{B_q}$ parameterizes the value of the
relevant hadronic matrix element and $S_0(x_t)$ gives the electroweak
loop contribution as a function of the top quark mass $m_t^2 = x_t
m_W^2$. Also the ratio $\Delta m_d / \Delta m_s$ is very useful since
it is simply related to $R_t$ via
\beq \label{SM-RDelMb}
{\Delta m_d \over \Delta m_s} \simeq 
 R_t^2 \, {\lambda^2 \over \xi^2} \cdot {m_{B_d} \over m_{B_s}} \,,
\eeq
where $\xi \equiv f_{B_s}/f_{B_d} \cdot \sqrt{B_{B_s}/B_{B_d}} \,.$

The measurement of the $\epsK$ parameter in combination with the
SM prediction determines a region between two hyperbolae in the
$\rho-\eta$ plane, described by~(see {\it e.g.}~\cite{BaBar,Buras2001}):
\beq \label{SM-eps}
\eta \left[(1-\rho) \eta_2 S_0(x_t) A^2 + P_c \right]
 A^2 B_K \simeq 0.226 \,.
\eeq
Here $\eta_2$ is a QCD correction factor and $P_c$ characterizes the
contributions with one or two intermediate charm-quarks.

Finally the recent results for the CP asymmetry in the $B^0_d \to
J/\psi \, K_S$ decay in~\eq{apsiKs} imply a further constraint in the
$\rho-\eta$ plane via the SM prediction
\beq \label{apsiKS}
a_{\psi K_S} = \sin2\beta \,,
\eeq 
where
\beq \label{betaSM}
\beta \equiv 
\arg\left(-{V_{cd} V_{cb}^* \over
            V_{td} V_{tb}^*}\right) \,.
\eeq
In terms of the Wolfenstein parameters we have
\beq \label{sin2betaSM}
\sin2\beta \simeq {2\eta(1-\rho) \over {\eta^2+(1-\rho)^2}} \,.
\eeq

Combining the above formulae and using the input parameters in Tab.~1
we obtain the allowed region for $(\rho,\eta)$ shown in
Fig.~\ref{rho-eta-SM}.  Note that we naively combine the
aforementioned constraints in order to determine the allowed region.
(For more sophisticated analyses see {\it e.g.}
refs.~\cite{WuZhou,AS,CKM2000,PS}.)

\section{Generic New Physics Models With Unitary Mixing Matrix}
\label{generic}

We focus our analysis on a large class of new physics (NP) models with
the following two features (c.f. refs~\cite{BEN,ENP,feature} and
references therein):
\begin{itemize}
\item[(i)] The $3\times3$ mixing matrix for the charged current weak
  interactions is unitary.
\item[(ii)] Tree-level decays are dominated by the SM contributions.
  
\end{itemize}
Note that assumption (i) holds in all SM extensions with three quark
generations. Assumption (ii) implies in particular that the phase of
the $B \to \psi K_S$ decay amplitude is given by the Standard Model
CKM phase, $\arg(V_{cb}V_{cs}^*)$.  Both assumptions are satisfied by
a wide class of models, like supersymmetry with $R$-parity, various
LRS models, models with minimal flavor violation (MFV) and several
multi-Higgs models. Since some of the models mentioned above have
additional features we investigate the following three scenarios:
\begin{itemize}
\item[(a)] {\it Only\/} assumptions (i) and (ii) are valid.
\item[(b)] Assumptions (i) and (ii) apply {\it and\/} 
  $\Delta m_d / \Delta m_s$ is unaffected by the presence of NP.
\item[(c)] Assumptions (i) and (ii) apply {\it and\/} 
  $\delta_{KM}$ is small.
\end{itemize}
Scenario (b) contains, for example, certain multi-Higgs models with
``flavor conservation'' and models with
MFV~\cite{Buras2001,BGGJS,MSSM,MFVold}.  Scenario (c) applies to
models of approximate CP symmetry and LRS models with spontaneous
CPV~\cite{Eck,Bar1,Bar2,Bar3,Bar,SP,Ball1,Ball2}.  To be specific we
take $|\tan(\delta_{KM})| < \tan(\delta^{max})=0.25$ for these
scenarios.

The NP effects relevant to our analysis modify the $B_d-\bar B_d$
mixing amplitude, $M_{12}-{i\over2}\Gamma_{12}$. Assumption (ii)
implies that the impact of NP on the absorptive part of this amplitude
is negligible:
\beq
\label{NP-Gamma}
\Gamma_{12} \approx \Gamma_{12}^{\rm SM} \,.
\eeq
The modification of $M_{12}$ can be parameterized as follows~(see
refs.~\cite{BEN,ENP,rdthd} and refs. therein):
\beq
\label{defrthe}
{M_{12} \over M_{12}^{\rm SM}} = r_d^2 \, e^{2i \theta_d} \,.
\eeq
The variables $r_d$ and $\theta_d$ relate directly to the experimental
observables.  The results for $\Delta m_{d,s}$ constrain $r_d^2$,
while the recent measurement of $a_{\psi K_S}$ restricts the value of
the phase $2\theta_d$.  However, to investigate a specific model it is
convenient to use different variables, $h$ and $\sigma$, and describe
the NP modifications to the amplitude $M_{12}={M_{12}^{\rm SM} +
  M_{12}^{\rm NP}}$ via~\cite{BEN,ENP,rdthd}:
\beq
\label{defrthe2}
{M_{12}^{\rm NP}\over M_{12}^{\rm SM}} = h \, e^{i\sigma} \,.
\eeq 
The variables $h$ and $\sigma$ are related to $r_d$ and $\theta_d$
through:
\beq \label{h-sigma}
r_d^2 \, e^{2i\theta_d}=1 + h \, e^{i\sigma} \,.
\eeq

Our goal is to determine the allowed parameter space for $h$ and
$\sigma$.  To this end we first constrain $r_d^2$ and $2\theta_d$
using the relevant experimental results on $R_u$, $\Delta m_{d,s}$,
$\eps_K$ and $\apks$. Then we show how to translate the ranges for
$r_d^2$ and $2\theta_d$ into the allowed region in the $h-\sigma$
plane.  All the relevant parameters we use in our analysis are
summarized in Tab.~1.  For most of these parameters we adopt the
values of ref.~\cite{CKM2000}.  For simplicity we scan over the ranges
indicated in Tab.~1 both for the theoretical and experimental
parameters.

\subsection{Constraining $r_d$ and $\theta_d$}
\label{rd-theta}

The parameter $r_d^2$ is constrained by the results for $\Delta
m_{d,s}$~\cite{BEN,ENP}.  In models that satisfy the assumptions~(i)
and~(ii) the expressions for $\Delta m_{d,s}$ in eq.~(\ref{SM-DelMb})
have to be modified as follows~\cite{BEN}:
\beq \label{NP-DelMb}
\Delta m_q = {G_F^2 \over 6 \pi^2} \eta_B m_{B_q} m_W^2 f^2_{B_q}
 B_{B_q} S_0(x_t) \, |V_{tq} V_{tb}^*|^2  \cdot r_q^2 \,,
\eeq
where $q=d,s$ and $r_s$ is the analog to $r_d$ for the $B_s$ system.
Moreover $\Delta m_{B_d}/\Delta m_{B_s}$ satisfies:
\beq \label{NP-RDelMb}
{\Delta m_{B_d} \over \Delta m_{B_s}} \simeq 
\lambda^2 R_t^2 \cdot \xi^{-2} {m_{B_d} \over m_{B_s}} 
 \left({r_d \over r_s}\right)^2 \,,
\eeq
which reduces to the SM expression in eq.~(\ref{SM-RDelMb}) when $r_d
= r_s$.  Note, that the constraint on $R_u$ given in eq.~(\ref{Rb}) is
unchanged by NP, since it is dominated by tree-level amplitudes.

Using the values in Tab.~1, from~\eq{NP-DelMb} it follows that:
\beq \label{rdDmd}
0.69 < && r_d \cdot R_t(\rho,\eta) < 1.03 \,,
\eeq
while~\eq{NP-RDelMb} implies that
\beq \label{rdDmd1}
{r_d \over r_s} R_t(\rho,\eta) \leq 0.99 \,.
\eeq

The allowed range for the phase $\theta_d$ is restricted by the
measurement of $a_{\psi K_S}$~\cite{BEN}.  
Due to the two-fold ambiguity of the $\sin$ function, 
the one-sigma result for $a_{\psi  K_S}$ in~\eq{apsiKs}
corresponds to two allowed intervals for
$2(\theta_d+\beta)$~\cite{BEN,ENP}:
\beq \label{betares}
19^\circ \lsim 2(\theta_d+\beta) \lsim 44^\circ \ \ \ \mbox{or} \ \ \ 
19^\circ \lsim 180^\circ-2(\theta_d+\beta) \lsim 44^\circ\,.
\eeq

We use the above results for two types of analysis: In our numerical
analysis we scan over the $\rho-\eta$ plane and calculate the allowed
intervals for $r_d$ according to \eq{rdDmd} [and~\eq{rdDmd1} in
scenario (b)] and $\theta_d$ from \eq{betares}
for all points ($\rho,\eta$) that fulfill the
constraints relevant to a particular scenario (c.f.
Fig.~\ref{rho-eta-a} for scenario (a), Fig.~\ref{rho-eta-b} for
scenario (b) and Fig.~\ref{rho-eta-c} for scenario (c), respectively).

To get better insight into the numerical results we have performed an
analytical calculation, where we use the maximal allowed region for
$r_d$ and $\theta_d$ that follow from the {\it global} constraints on
$R_t$ and $\beta$. Such an approach is particularly useful, because
the analytical constraints can easily be updated using the forthcoming
experimental data. Moreover the system of constraints may be extended
for different classes of NP models.  For scenario (a) we get (see
appendix~\ref{global} for details):
\beq \label{rd-theta-a}
r_d \in [0.5,~1.8] \andthis 2\theta_d \in [-32^\circ,~212^\circ] \,.
\eeq
Note that the interval for $2\theta_d$ in \eq{rd-theta-a} is
independent of the upper bound on $\apks$ as long as
$(2\theta_d)_{max_1}>(2\theta_d)^{min_2}$. With the upper bound on
$R_u$ in \eq{Rb} two distinct intervals for $2\theta_d$ only appear if
$(\apks)_{max} < 0.63$. This explains why the global constraint on
$2\theta_d$ we obtain from the measurement of $\apks$ in \eq{apsiKs}
almost coincides with the one in ref.~\cite{BEN}, even though our
upper bound on $\apks$ is much more stringent.

For scenario (b) the lower bound on $r_d$ is somewhat stronger, since
from \eq{rdDmd1} it follows that $R_t(\rho,\eta) \leq 0.99$, which
gives $r_d > 0.7$. In scenario (c) we have
\beq 
r_d \in [0.5,~0.8] \withthis 
2\theta_d \in [12^\circ,~51^\circ] \cup [129^\circ,~168^\circ] \,, 
\eeq
for the allowed regions with $\rho<0$ and
\beq \label{rd-theta-c} 
r_d \in [1.0,~1.8] \withthis 
2\theta_d \in 
[-1^\circ,~64^\circ] \cup [116^\circ,~181^\circ] \,, 
\eeq
for the allowed regions with $\rho>0$ (c.f. Fig.~\ref{rho-eta-c}).

\subsection{Constraining $h$ and $\sigma$}
\label{h-sig}

Our first observation is that from the maximal value for $r_d$ one
obtains an upper bound on $h$. \Eq{h-sigma} implies that
\beq \label{h-rd-theta}
h(r_d,\theta_d) = \sqrt{r_d^4 - 2r^2_d \cos2\theta_d +1} 
               \le (r_d^{max})^2 + 1 \approx 4.24 \,,
\eeq
where the upper bound arises from setting $\cos2\theta_d=-1$ and
$r_d=r_d^{max}$.  Still, one can obtain much better bounds using the
correlation between $h$ and $\sigma$. Here we only sketch how to
translate the constraints one obtains for $r_d^2$ and $2\theta_d$ into
those for $h$ and $\sigma$. The details can be found in
appendix~\ref{h-sig-details}.  The basic idea is simple. We solve
\eq{h-sigma} for $h$, once eliminating $\theta_d$,
\beq \label{h-sigma-rd}
h^\pm(\sigma,r_d) \equiv 
 -\cos\sigma \pm \sqrt{\cos^2\sigma+r_d^4-1} \,,
\eeq
and once eliminating $r_d$,
\beq \label{h-sigma-theta}
h^\pm(\sigma,\theta_d) = 
{\sin2\theta_d \, \sin(2\theta_d \pm \sigma)
 \over \sin^2\sigma -\sin^2 2\theta_d} \,.
\eeq
Using the above equations the ranges for $r_d$ and $\theta_d$ can be
translated into a region in the $h-\sigma$ plane. The overlap of the
two areas from \eq{h-sigma-rd} and \eq{h-sigma-theta} corresponds to
the allowed region in the $h-\sigma$ parameter space. While this
method is straightforward one has to treat carefully the discrete
ambiguities of the trigonometric functions (see appendix).

Using \eq{h-sigma-rd} and \eq{h-sigma-theta} the global constraints on
$r_d$ and $\theta_d$ in \eqs{rd-theta-a}--(\ref{rd-theta-c}) can be
readily translated into the allowed region for $h$ and $\sigma$.
Combining all the resulting intervals one obtains the light areas,
whose boundaries are given by $h^\pm(\sigma, r_d)$ and $h^\pm(\sigma,
\theta_d)$ with $r_d$ and $\theta_d$ at the edges of their allowed
intervals (see appendix~\ref{h-sig-details} for more details).

In order to obtain the exact allowed region in the $h-\sigma$ plane we
have performed a numerical analysis proceeding as follows: Rather than
using the global constraints on $r_d$ and $\theta_d$ corresponding to
the entire allowed region in the $\rho-\eta$ plane, we compute the
possible ranges for $r_d$ and $\theta_d$ for {\it each} permissible
value of $\rho$ and $\eta$. That is we scan over the $\rho-\eta$ plane
and for all points ($\rho,\eta$) that fulfill the constraints relevant
to a particular scenario, we calculate the allowed intervals for $r_d$
according to \eq{rdDmd} [and~\eq{rdDmd1} in scenario (b)] and
$\theta_d$ from \eq{betares}. The resulting
intervals are then translated into the corresponding region in the
$h-\sigma$ plane, just as we did in the ``global'' analysis, but for
each point ($\rho,\eta$) separately.  Combining all the resulting
intervals one obtains the allowed regions indicated as the dark areas
in Fig.~\ref{h-sigma-a}, Fig.~\ref{h-sigma-b} and
Fig.~\ref{h-sigma-c}.  They are contained within the light areas
resulting from the analytic boundaries from the global constraints.
Even though in general the numerical results are not much more
constraining than the global ones, there exist some regions where the
numerical constraints are significantly stronger than the analytical
boundaries.  For example, there appear ``holes'' in the dark gray
regions of Fig.~\ref{h-sigma-a} and Fig.~\ref{h-sigma-b} that
correspond to the excluded region within the $R_u$ annulus, which has
been ignored in ``global'' analysis (light gray regions). Even more
important, the lower bound on $h$ in scenario (c) (see
Fig.~\ref{h-sigma-c}) from the numerical analysis turns out to be
significantly stronger than in the global analysis, which is
marginally consistent with $h=0$.

\subsection{Summary of Results}

\begin{itemize}
\item[(a)] Generic models: All models that satisfy the assumptions (i)
  and (ii) must respect the condition in~\eq{Rb}, which corresponds to
  an annulus around the origin in the $\rho-\eta$ plane as shown in
  Fig.~\ref{rho-eta-a}.  The resulting allowed region in the
  $h-\sigma$ plane is presented in Fig.~\ref{h-sigma-a}. The
  admissible range for $h$ is given by:
  \beq \label{h-range-a}
  h \in [0,~4.2] \,.
  \eeq
    
\item[(b)] Models where $\Delta m^{SM}_d / \Delta m^{SM}_s \approx
  \Delta m_d / \Delta m_s$: The relevant region in the $\rho-\eta$
  plane is shown in Fig.~\ref{rho-eta-b} and the resulting allowed
  region in the $h-\sigma$ plane is presented in Fig.~\ref{h-sigma-b}.
  The admissible range for $h$ is the same as in \eq{h-range-a}, but
  the excluded area in the $h-\sigma$ plane is somewhat larger.
  
\item[(c)] Models in which $\delta^{KM}$ is significantly smaller than
  unity: The relevant region in the $\rho-\eta$ plane for this class
  of models is shown in Fig.~\ref{rho-eta-c} and the resulting allowed
  region in the $h-\sigma$ plane is presented in Fig.~\ref{h-sigma-c}.
  The excluded region is significantly smaller than in scenario (a)
  and (b). The admissible range for $h$ and $\sigma$ are given by:
  \beq \label{h-approxCP}
  h \in [0.14,~4.18] \andthis \sigma \in [0^\circ,~180^\circ] \,.
  \eeq
  Unlike for scenarios (a) and (b) in this case $h$ has
  also a non-trivial lower bound and $\sigma$ has an upper bound.
  Consequently NP contributions to the $B-\bar B$ mixing amplitude 
  are required in this scenario. 

\end{itemize}

\section{Specific New Physics Models}

In this section we focus on two specific New Physics models that
belong to the general class of models discussed above. We discuss
first models with minimal flavor violation (MFV) and subsequently we
present some new results relevant to LRS models with spontaneous CPV.

\subsection{Minimal Flavor Violation}

Models with MFV comprise the SM and those extensions of the SM in
which all flavor changing interactions are described by the CKM
matrix. These models do not introduce any new operators implying the
absence of any new CP violating phases beyond the KM phase. The only
impact of MFV models are modifications of the Wilson coefficients of
the SM operators due to additional contributions from diagrams
involving new internal particles~\cite{Buras2001,BGGJS,MSSM,MFVold}.

\subsubsection{Constraints from $R_u$, $\Delta m_{d,s}$ and $\eps_K$}

In the framework of this analysis the only relevant modifications of
the SM predictions concern the mass difference $\Delta m_d$ of the
$B_d$ system and the parameter $\epsK$ describing CP violation in
$K^0-\bar K^0$ mixing.  It turns out that for both parameters the new
physics charm and charm-top contributions are negligible and that
additional contributions to the SM box diagram with top quark
exchanges can be described by one single parameter $F_{tt}$.

This parameter effectively replaces the relevant Inami-Lim function
$S_0(x_t)$ appearing in eqs.~(\ref{SM-DelMb}) and~(\ref{SM-eps}),
which have to be substituted by~\cite{BGGJS}
\beq \label{MFV-DelMb}
\Delta m_d = {G_F^2 \over 6 \pi^2} \eta_B m_{B_d} m_W^2 f^2_{B_d}
 B_{B_d} F_{tt} \cdot |V_{td} V_{tb}^*|^2 \,.
\eeq
and
\beq \label{MFV-eps}
\eta \left[(1-\rho) \eta_2 F_{tt} A^2 + P_c \right]
 A^2 B_K \simeq 0.226 \,,
\eeq
respectively. The constraint on the unitarity triangle from $R_u$
[c.f. \eq{Rb}] remains unaffected in MFV models, because it is
dominated by tree-level contributions. Furthermore, due to the absence
of any new phases in the $B-\bar B$ mixing amplitude the SM prediction
for $\apks$ [c.f. \eqs{apsiKS}-(\ref{sin2betaSM})] is unchanged.
Finally, $\Delta m_d / \Delta m_s$ in eq.~(\ref{SM-RDelMb}) remains
unaffected since the NP contributions cancel each other in the ratio.

The important point we want to make is that for any given set of the
parameters $R_u, A, B_K, P_c$ and
\beq
\Delta_f \equiv {\Delta m_d \over B_{B_d} f_B^2 \lambda^6} \,
\eeq
one can obtain exact expressions for $\rho$, $\eta$ and $F_{tt}$.
Combining eqs.~(\ref{Rb}), (\ref{MFV-DelMb}) and~(\ref{MFV-eps})
results into a quartic equation that can be solved analytically. For
example, for $\rho$ one obtains
\beq
(R_u^2-\rho^2) \, 
\left[(1-\rho) C_\Delta + C_c (1-2\rho+R_u^2)\right]^2 = 
\left[C_\varepsilon (1-2\rho+R_u^2)\right]^2 \,,
\eeq
where
\beq
C_\Delta \equiv {6 \pi^2 \Delta_f \over G_F^2 \eta_B m_{B_d} M_W^2
A^2} \,, \ \ \ \ \
C_\varepsilon \equiv {0.226 \over A^4 B_K \eta_2} \,, \andthis 
C_c \equiv {P_c \over A^2 \eta_2} \,. 
\eeq
We have used the analytic expressions to obtain the allowed regions
for $\rho$, $\eta$ and $F_{tt}$ by scanning over the intervals for
$R_u, A, B_K, P_c$ and $\Delta_f$ in Tab.~2.  {\it A priori} up to
four solutions are possible, but for the present ranges of the
parameters $R_u, A, B_K, P_c$ and $\Delta_f$ only two solutions are
real (see Fig.~\ref{MFV-AR1} and Fig.~\ref{MFV-AR2}).  Adding the
constraint from $\Delta m_d / \Delta m_s$ in eq.~(\ref{SM-RDelMb})
practically the entire solution shown in Fig.~\ref{MFV-AR1} is ruled
out and only the second solution in Fig.~\ref{MFV-AR2} remains viable.
We find that $F_{tt}$ is restricted to the interval
\beq \label{Ftt-range}
F_{tt} \in [1.2,~5.7] \,,
\eeq
which includes the SM value $F_{tt}^{SM} \equiv S_0(x_t) \simeq 2.46
\, (m_t/170~{\rm GeV})^{1.52}$~\cite{Buras2001}.

The value of $\sin 2\beta$ within models of MFV has obtained
considerable attention recently~\cite{BurasBuras,Buras2001,WuZhou}. In
particular it was shown in ref.~\cite{BurasBuras} that there exists a
lower bound on $\sin 2\beta$ in these models.  The existence of this
bound is a non-trivial result of the correlation between $\Delta m_d$
and $\eps_K$ which holds in models of MFV~\cite{BurasBuras}.  Using
the analytic solutions for $\rho$, $\eta$ and $F_{tt}$ we have
investigated the behavior of $\sin 2\beta$ as a function of $R_u, A,
B_K, P_c$ and $\Delta_f$.  We find that it takes its minimal value,
$(\sin 2\beta)_{min}$, when $R_u$ is minimal and all the other
parameters are maximal.  While the size of the exact expressions we
obtained for $\rho$, $\eta$, $F_{tt}$ and $\sin 2\beta$ preclude its
presentation here, one can use the results to expand these parameters
around a particular value. For example a linear expansion of $\sin
2\beta$ around $R_u^{min}, A^{max}, B_K^{max}, P_c^{max}$ and
$\Delta_f^{max}$ gives:
\beqs \label{s2b-min-exp}
\sin 2\beta &\simeq& 0.52 + 0.23 \, (R_u-0.34) - 1.09 \, (A-0.86)
- 0.47 \, (B_K-1.00)  \nonumber \\
&~&  - 0.31 \, (P_c-0.36) 
- 3.43 \times 10^{-6} \, 
\left({\Delta_f \over {\rm ps}^{-1}\,{\rm GeV}^{-2}}-1.04 \times 10^5
\right) \,.
\eeqs
This expansion provides a good approximation to the exact expression
for $\sin 2\beta$ within most of the allowed parameter space for $R_u,
A, B_K, P_c$ and $\Delta_f$. Moreover we find that for reasonable
changes in the boundaries of the respective intervals, $\sin 2\beta$
remains minimal when $R_u$ is minimal and all the other parameters are
maximal.  Thus one can use eq.~(\ref{s2b-min-exp}) to recalculate
$(\sin 2\beta)_{min}$ for somewhat different values of $R_u^{min},
A^{max}, B_K^{max}, P_c^{max}$ and $\Delta_f^{max}$.  For example
using the more conservative values of ref.~\cite{Buras2001} one
recovers the corresponding lower bound on $\sin 2\beta$ which is
somewhat weaker than our result,
\beq \label{s2b-min}
(\sin 2\beta)_{min} = 0.52 \,.
\eeq
We note that the upper bound on $\sin 2\beta$,
\beq
(\sin 2\beta)_{max} = 2 R_u^{max} \sqrt{1-(R_u^{max})^2} = 0.78 \,,
\eeq
is determined by the tangent to the $R_u$ annulus as in the
SM~\cite{BOL}.

Conversely, the range of $\sin 2\beta$ due to the measured value of
$\apks$ restricts the parameter $F_{tt}$. Remarkably, taking the
one-sigma range in~\eq{apsiKs}, only the lower bound on $F_{tt}$ is
slightly improved to $F_{tt}^{min}=1.3$. The reason is evident from
Fig.~\ref{MFV-AR2}: Even though the upper bound on $\sin 2\beta$
removes a substantial part of the allowed region, the remaining
solution (below the upper gray line) still extends over a wide range
for the parameter $F_{tt}$ (as indicated by the gray-scale).

Finally we note that the allowed interval for $F_{tt}$ is consistent
with the result obtained in the generic framework discussed in
section~\ref{generic}. Indeed the ranges for $h$ at $\sigma=0$ or
$\sigma=\pi$ in Fig.~\ref{h-sigma-b} include the values for $h^{MFV}$
defined via
\beq
h^{MFV} e^{i\sigma^{MFV}} \equiv (F_{tt}/F_{tt}^{SM} - 1) \in [-0.48,~1.51] \,,
\eeq
where -- due to the absence of new phases in MFV models -- either
$\sigma^{MFV} = 0$ if $F_{tt} \ge F_{tt}^{SM}$ or $\sigma^{MFV} = \pi$
if $F_{tt}<F_{tt}^{SM}$.

To conclude, models of MFV provide a consistent explanation of all the
relevant experimental data. However, due to the $\Delta m_d / \Delta
m_s$ constraint the allowed region in the $\rho-\eta$ plane is similar
to the one of the SM and $F_{tt}$ is close to $F_{tt}^{SM}$.
Therefore with the set of observables studied in this section it is
very difficult, if not impossible, to distinguish MFV models beyond
the SM.  Therefore, we turn now briefly to a second set of observables
which might help to disentangle such models from the SM.

\subsubsection{Constraints from $Y \to X \nu \bar\nu$}

Let us consider the decays $K^+ \to \pi^+ \nu \bar\nu$ and $B \to X_s
\nu \bar\nu$ within the MFV models. They are both related to the
generalized Inami-Lim~\cite{IL} function, $X_{\nu\bar\nu}$~\cite{BGGJS,MSSM}.
The theoretical prediction for $K^+ \to \pi^+ \nu \bar\nu$ is given
by~\cite{Buras2001,BGGJS}:
\beq \label{bkpn}
{\cal B}_K \equiv \BR(K^+ \to \pi^+ \nu \bar\nu)/\kappa_+ \approx
\left\{ 
\left(A^2 \eta X_{\nu\bar\nu} \right)^2 +
\left[P_0(X)+A^2 (1-\rho) X_{\nu\bar\nu} \right]^2\right\} \,,
\eeq
where the values of $P_0(X)$ and $\kappa_+$ are given in Tab.~1.

Using \eq{bkpn} from the experimental value of $\BR(K^+
\to\pi^+\nu\bar\nu)$ (given in Tab.~1) and the value of $\kappa^+$ we
get:
\beq \label{bkplcon}
0.7 = {\cal B}^{min}_K < 
\left(A^2 X_{\nu\bar\nu}R_t\right)^2 + 2 P_0(X)
    A^2 \, X_{\nu\bar\nu} \, (1-\rho) + P_0(X)^2 <
{\cal B}^{max}_K = 11.9 \,,
\eeq
where ${\cal B}^{min}_K$ (${\cal B}^{max}_K$) denotes the minimal
(maximal) value of ${\cal B}_K$.  From \eq{bkplcon} it follows that
$X_{\nu\bar\nu} \in [X^{min}_+,~X^{max}_+]$ or $X_{\nu\bar\nu} \in
[X^{max}_-,~X^{min}_-]$, where
\beqs \label{Xsolution}
X^{min}_\pm &\equiv& 
{1 \over A^2 R_t^2} \left[P_0(X)(\rho-1) \pm 
 \sqrt{R_t^2 {\cal B}_K^{min}-\eta^2 P_0(X)}\right] \,,\\
X^{max}_\pm &\equiv& 
{1 \over A^2 R_t^2} \left[P_0(X)(\rho-1) \pm 
 \sqrt{R_t^2 {\cal B}_K^{max}-\eta^2 P_0(X)}\right] \,.
\eeqs 
A scan of the allowed $\rho-\eta$ region in Fig.~\ref{MFV-AR2}
(including the $\apks$ constraint) and the intervals for $A$ and
$P_0(X)$ yields the following two ranges for
$\left(X_{\nu\bar\nu}\right)$:
\beq \label{X_tppl}
0.5  < X_{\nu\bar\nu} <  7.8 
\orthis
-9.8 < X_{\nu\bar\nu} < -1.6 \,.
\eeq
Note that the uncertainty related to the measurement of~$\BR(K^+
\to\pi^+\nu\bar\nu)$ (given in Tab.~1) is based on observation
of a single event.
Therefore, 
the ranges given in \eq{X_tppl} correspond to a confidence level of somewhat
less than 67 \%.

Within MFV models the decay $B\to X_s \nu \bar\nu$ is also related to
the parameter $X_{\nu\bar\nu}$~\cite{Buras2001,BGGJS}. The prediction
for the branching ratio,
\beq \label{bbxnn}
\BR(B\to X_s\nu\bar\nu) \approx 
1.5 \cdot 10^{-5} \cdot X_{\nu\bar\nu}^2 \,,
\eeq
in combination with the experimental bound (given in Tab.~1):
\beq \label{expBXs}
\BR(B\to X_s\nu\bar\nu) < 6.4 \cdot 10^{-4} \,,
\eeq
provides a second constraint on $X_{\nu\bar\nu}$: 
\beq \label{BXscon}
\left|X_{\nu\bar\nu}\right| \lsim 6.6 \,,
\eeq
which excludes some of the allowed range given in \eq{X_tppl}.
Combining the two ranges yields:
\beq \label{x_tcomb}
 0.5 < X_{\nu\bar\nu} < 6.6 
\orthis
-6.6 < X_{\nu\bar\nu} < -1.6 \,.
\eeq

We note that in principle also the decay $B\to X_d \nu\bar\nu$ could
provide a constraint on $X_{\nu\bar\nu}$, but since it is suppressed
by $V_{td}/V_{ts}$ with respect to $B\to X_s \nu\bar\nu$ its branching
ratio is decreased by roughly an order of magnitude and therefore at
present less useful.

Currently the constraints are too weak to rule out one of the two
intervals, but the forth-coming data on $K$ and $B$ decays with final
neutrinos will eventually allow to pin down the value of
$X_{\nu\bar\nu}$ and maybe provide evidence for MFV beyond the SM if
$X_{\nu\bar\nu}$ turns out to be different from its SM value
$X_{\nu\bar\nu}^{SM}=X(x_t) \approx 1.5$~\cite{Buras2001}. In this
context also the observable $(\eps'/\eps)_K$ could be useful, since it
depends on a related parameter $X_{u \bar u}$, which in some models
roughly agrees with $X_{\nu\bar\nu}$~(compare for example the values
for $X_{u \bar u}$ and $X_{\nu\bar\nu}$ in ref.~\cite{MSSM}).

\subsection{LRS Model with Spontaneous CP Violation}

The spontaneously broken LRS (SBLR) model has been studied in detail
in refs.~\cite{Eck,Bar1,Bar2,Bar3,Bar,SP,Ball1,Ball2} (see also
references therein).  Here we focus on the impact of the measurement
of $\apks$ in \eq{apsiKs} on this model.  The important feature of the
SBLR model is that essentially all the phases and therefore the CPV
observables depend on one parameter. In a certain phase
convention~\cite{Bar1,Bar2,Bar3,Bar} this parameter is written as
$r\sin\alpha$ with $|r|\lsim m_b/m_t$ (see also
refs.~\cite{Eck,Ball1}).  Due to this single parameter the model is
very predictive.

The problem is that part of the model has not yet been solved
analytically. In particular analytic expressions for the phases of the
left and right CKM-like mixing matrices exist only within the ``small
phase approximation''~\cite{Eck,Bar1,Bar2,Bar3,Bar}, which is valid
for $r\sin\alpha \lsim 0.01$.  Nevertheless a very thorough numerical
analysis, beyond the small phase approximation, has been performed by
Ball {\it et. al.} in ref.~\cite{Ball1}, calculating the predictions
for several CPV observables and discussing the limitation of the small
phase approximation. In particular it was found that within the LRSB
model the measurement of $\epsK$ in \eq{epsK} and other observables
implies
\beq \label{apks-LRSB}
(\apks)_{LRSB} \lsim 0.1 \,,
\eeq
which is inconsistent with the combined measurement in \eq{apsiKs}.
It is important to note, however, that the analysis in
ref.~\cite{Ball1} only used the central values for various input
parameters which are subject to theoretical and experimental
uncertainties.  Therefore we find it timely to reinvestigate the
prediction in \eq{apks-LRSB} including these uncertainties and using
the most recent values for the central values.

\subsubsection{The Analysis}

We restrict our analysis to the regime in which the small phase
approximation is valid, which allows us to use the analytic expression
for various observables of the $K$ and $B$ system.

We scan over the entire parameter space of the relevant input
parameters according to their uncertainties in order to find the
subspace consistent with all the measured observables.  In particular,
we test all the 32 different sign combinations of the quark masses,
and allow the CKM phase $\delta$ to be centered both around 0 and
$\pi$ (c.f. discussion in ref.~\cite{Branco}) which yields altogether
64 different signatures.  For each permissible subspace we calculate
the corresponding predicted range for $\apks$.

Most of the analytic expression in the small phase approximation have
been calculated in refs.~\cite{Bar1,Bar2,Bar3}.  For notations and
phase conventions we refer the reader to refs.~\cite{Bar2,Bar3}. (Note
that the parametrization of CKM matrix used in ref.~\cite{Bar1} is the
Kobayashi-Maskawa convention~\cite{CKM} and not the Chau-Keung
convention~\cite{CK}, which is commonly used.) All the parameters
relevant to our analysis can be found in Tab.~2.

For the kaon system we consider the observables $\Delta m_K$, $\epsK$
and $(\eps'/\eps)_K$.  The effective Lagrangian ${\cal L}^{\Delta
  S=2}$, that describes the $\Delta S=2$ processes, is given in
eq.~(19) of ref.~\cite{Bar2}.  The relevant phases of the left and
right mixing matrices, $\delta_1$, $\delta_2$, $\gamma$ and $\delta$
have been calculated in ref.~\cite{Bar1}. We use the following three
constraints:

\begin{itemize}
\item[(i)] $\Delta m_K$: Due to the lack of reliable predictions for
  the long distance contributions we require that the short distance
  contribution should at most saturate $\Delta m_K$, i.e.
\beq \label{delmK}
2 |M_{12}^{K}| < \Delta m_K^{\rm  exp} \,,
\eeq
where $M_{12}^K = \langle K^0 | {\cal L}^{\Delta s=2}| \bar{K}^0
\rangle / (2 m_K) $.

\item[(ii)] $\eps_K$: We use the upper and lower bound given in
  ref.~\cite{Ball1}:
\beq \label{tilth}
6.375 \cdot 10^{-3} - 0.01 \cdot 
 \left(\frac{1\,{\rm TeV}}{M_2}\right)^2
 < \widetilde{\theta}_{M} < 
6.523 \cdot 10^{-3} + 0.01 \cdot 
 \left(\frac{1\,{\rm TeV}}{M_2}\right)^2 \,,
\eeq
where, in our case, $\widetilde{\theta}_{M} \approx \left|\frac{2 \,
{\rm Re}\,M_{12}}{\Delta m_K}\right| \sin\left(\arg M_{12}\right)$.

\item[(iii)] $\Re(\eps'/\eps)_K$: Due to the large hadronic
  uncertainties in the calculation of $\eps'_K$ we only demand that
  the sign predicted by the SBLR model is positive in order to be
  consistent~\cite{Ball1} with recent measurements~\cite{eps-prime}.
  Thus, for the computation of $\Re(\eps'/\eps)_K$ it is sufficient to
  use the rough approximation given in eqs.~(27) and (28) of
  ref.~\cite{Bar1}.
\end{itemize}

For the $B$ system we consider $\Delta m_q$ ($q=d,s$) and $\apks$.
The value of $h_q^{\rm LR} \equiv {{M^q_{12}}^{\rm LR} \over
{M^q_{12}}^{\rm SM}}$, where $M^q_{12}=\langle B^0_q | {\cal
L}^{\Delta b=2}| \bar{B}^0_q\rangle/(2 m_q)$, has been calculated in
ref.~\cite{Ball1}:
\beq \label{hLR}
h^{\rm LR}_q \approx h^{\rm LR} \equiv {B_B^S (m_b) \over B_B (m_b)}
 \left[1.7 \, \left(0.051 + 0.26 \ln{M_2\over1.6~{\rm TeV}}\right)
   \left({1.6~{\rm TeV}\over M_2}\right)^2
   +\left({7~{\rm TeV}\over M_H}\right)^2\right] \,.
\eeq
For $\sigma_q^{\rm LR}$ an exact
expression as a function of the fundamental parameters does not exist,
so we use the small phase approximations~\cite{Bar3}:
\beq \label{sinsigd}
\sin\sigma_d^{LR} &\approx& \eta_d\eta_b \, r \sin\alpha
 \left[\left( {\sin^2 t_1\over\eta_d m_d} 
  +{2\over \eta_s m_s}\right)\eta_c m_c 
  +{\eta_t m_t\over\eta_c  m_c} \right] \,,
\\ \label{sinsigs}
\sin\sigma_s^{LR} &\approx& \eta_s\eta_b \, r \sin\alpha
 \left({\eta_c m_c\over \eta_s m_s} 
  +{\eta_t m_t\over \eta_b m_b}\right) \,,
\eeq 
where $t_1$, $t_2$ and $t_3$ denote the three angles of the CKM matrix
in the Chau-Keung convention~\cite{CK} and $\eta_i = \pm 1$
parameterizes the sign associated with the quark mass $m_i$. For our
analysis we use the following three constraints:
\begin{itemize}
\item[(i)] $\Delta m_d$: \Eq{rdDmd}, \eq{hLR} and \eq{sinsigd} yield
  the constraint:
\beq \label{dmdcon}
0.48 < \left| (V_{tb}^L V_{td}^{L*})^2 (1+h^{\rm LR} \, 
       e^{i\sigma^{\rm LR}_d}) \right| < 1.05 \,,
\eeq
where $V_{ij}^L$ is the mixing matrix of the left handed current of
the weak interaction.

\item[(ii)] $\Delta m_d \over \Delta m_s$: \Eq{rdDmd1}, \eq{hLR},
\eq{sinsigd}, and \eq{sinsigs} yield the constraint:
\beq \label{dmscon2}
\left|{(V_{td}^{L*})^2 (1+ h^{\rm LR}\, e^{i\sigma^{\rm LR}_d}) \over
       (V_{ts}^{L*})^2 (1+h^{\rm LR} \, e^{i\sigma^{\rm LR}_s})}
     \right| < 0.98 \,.
\eeq
\item[(iii)] $\apks$: In the SBLR model $\apks$ is given by~\cite{Ball1}:
\beq
\apks \simeq \sin \left[ 2 \beta_{\rm CKM}  +
 \arg\left(1 + h^{LR} e^{i \sigma_d^{LR}}\right)\right]\,,
\eeq
where
\beq 
\beta_{\rm CKM} \equiv 
\arg\left(-{V_{cd}^L V_{cb}^{L*} \over
            V_{td}^L V_{tb}^{L*}} \right) \,.
\eeq 
We require that $\apks$ is positive.
\end{itemize}

\subsubsection{Results}
We find positive values for $\apks$ that are consistent with all the
experimental constraints for several signatures.  The allowed range
for $\apks$ that results from scanning over the allowed ranges of the
relevant parameters in Tab.~2 is:
\beq \label{rangebeta}
\apks \in [0,~0.3] \,.
\eeq
In general, large values for $\apks$ occur for large values of
$r\sin\alpha$ for which the small phase approximation is less
reliable. Still, even for the rather large values of $r\sin\alpha$ one
still obtains a rough estimate for the true value of $\apks$.
Therefore the fact that the range given in \eq{rangebeta} is
marginally consistent with the world average of $\apks$ in \eq{apsiKs}
gives strong indication that the inclusion of the uncertainties of the
relevant parameters relaxes the upper bound on $\apks$ in
\eq{apks-LRSB} enough to make an exclusion of the LRSB model premature
at this stage.

\section{Conclusions}
\label{conclusion}

The CKM picture of the Standard Model (SM) is currently being tested
with unprecedented accuracy in various experiments. In particular the
$B$ factories BELLE and BaBar provide highly interesting
results. There is great hope that those or future experiments will
reveal inconsistencies that indicate contributions to flavor physics
from New Physics.

In this paper we have studied extensions of the Standard Model where
the charged current weak interactions are governed by the CKM matrix
and where all tree-level decays are dominated by the Standard Model
contributions.  We have constrained both analytically and numerically
the ratio $h$ and the phase difference $\sigma$ between the New
Physics and the Standard Model contributions to the mixing amplitude
of the neutral $B$ system using the experimental results on $R_u$,
$\Delta m_{d,s}$, $\eps_K$ and $\apks$.  For generic models we find
that $h<4.2$ and the allowed region in the $h-\sigma$ plane shown in
Fig.~\ref{h-sigma-a}.  Models where the ratio $\Delta m_d/\Delta m_s$
takes the Standard Model value are only slightly more constrained than
the most general scenario of our framework (see Fig.~\ref{h-sigma-b}).
However imposing a small CP violating phase significantly reduces the
allowed parameter space for $h$ ($h \in [0.14,~4.18]$) and 
$\sigma$ ($\sigma \in [0^\circ,~180^\circ]$), 
see Fig.~\ref{h-sigma-c}, which requires NP
contributions to the $B-\bar B$ mixing amplitude.

We have presented some new results for models with minimal flavor
violation, pointing out that the three parameters $\rho$, $\eta$ and
$F_{tt}$ can be obtained analytically from the experimental results on
$R_u$, $\Delta m_d$ and $\eps_K$ for any given set of rather
well-determined parameters. Using the exact expressions we have
updated the allowed interval of $F_{tt}$ ($F_{tt} \in [1.3,~5.7]$) as well as
$\sin 2\beta$ ($\sin 2\beta \in [0.52,~0.78]$) 
[for which we provide an expansion in
terms of those parameters in \eq{s2b-min-exp}].  Taking into account
the analytical relation between $\rho$, $\eta$ and $F_{tt}$ as well as
the constraints from $\apks$ and $\Delta m_d / \Delta m_s$ improves
significantly previous results~\cite{WuZhou}.

We also consider the spontaneously broken left-right symmetric model
and perform a numerical analysis using the ``small phase
approximation'' in order to probe the viability of this model in view
of the recent results for $\apks$ and other observables. We find that
the inclusion of the uncertainties of various input parameters relaxes
the upper bound on $\apks$ significantly and conclude that at present
the model is still viable.

\acknowledgements
 
We thank Y. Nir and Y. Grossman for helpful discussions and comments
on the manuscript.

\newpage

\appendix

\section{}

\subsection{Global constraints on $r_d$ and $\theta_d$}
\label{global}

To each of the three scenarios (a), (b) and (c) corresponds a
different allowed area in the $\rho-\eta$ plane (shown in
Fig.~\ref{rho-eta-a}, Fig~\ref{rho-eta-b} and Fig~\ref{rho-eta-c},
respectively). Here we compute the {\it global} constraints on $R_t$
and $\beta$ for each scenario.  From \eqs{Rb-interval} and
(\ref{rdDmd1}) it follows that in scenario (a) the allowed interval
for $R_t(\rho,\eta)$ [defined in \eq{Rt}] is given by
\beq \label{RtRange-a}
0.57 = 1-R_u^{max} < R_t(\rho,\eta) < 1+R_u^{max} = 1.43 \,.
\eeq
In scenario (b) the upper bound on $R_t$ is significantly stronger,
namely $R_t<0.99$ due to \eq{rdDmd1}. In scenario (c) there are two
ranges, corresponding to the two allowed regions in the $\rho-\eta$
plane in Fig.~\ref{rho-eta-c}. The upper bound on $\delta_{KM}$
implies that we have to exclude the interval
\beq \label{RtRange-c}
0.68 = \sqrt{(1-l_\delta)^2 + h_\delta^2} 
< R_t(\rho,\eta) <
\sqrt{(1+l_\delta)^2 + h_\delta^2} = 1.33 \,,
\eeq
where $l_\delta \equiv R_u^{min} \cos\delta^{max}$ and $h_\delta
\equiv R_u^{min} \sin\delta^{max}$, from the range in \eq{RtRange-a}
in this scenario.

In scenario (a) and (b) the angle $\beta(\rho,\eta)$ [defined in
\eq{betaSM}] is restricted by the maximal value of $R_u$:
\beq \label{beta-bound}
|\beta(\rho,\eta)| < \arcsin(R_u^{max}) = 25^\circ \,.
\eeq
In scenario (c) we have
\beq \label{beta-bound-c}
|\beta(\rho,\eta)| < 
\arctan\left({R_u^{max} \sin \delta^{max} \over 
             1 \pm R_u^{max} \cos \delta^{max}}\right) \simeq 
\left\{\matrix{3.5^\circ \cr 10^\circ} \right. \,,
\eeq
where the plus sign refers to the region with negative $\rho$ and the
minus sign to the region with positive $\rho$.  Using the above
equations one obtains the allowed intervals for $r_d$ and $\theta_d$
quoted in section~\ref{rd-theta} in
\eqs{rd-theta-a}--(\ref{rd-theta-c}).

\subsection{Analytic Boundaries in the $\h-\sigma$ Plane}
\label{h-sig-details}

In this appendix we describe in detail how the allowed intervals for
$r_d$ and $\theta_d$ are translated into the allowed region in the
$h-\sigma$ plane.  To be specific we will use the global extrema for
$r_d$ and $\theta_d$ in section~\ref{rd-theta}
[c.f. \eqs{rd-theta-a}--(\ref{rd-theta-c})].  Note, however, that
essentially the same equations apply for the numerical analysis, where
we consider each viable point in the $\rho-\eta$ plane separately.

\subsubsection{The $r_d$ constraint}
In order to obtain an equation that is independent of $\theta_d$ we
take the absolute square of \eq{h-sigma}:
\beq \label{rd-squared}
r_d^2 = \sqrt{h^2 + 2h\cos\sigma + 1} \,.
\eeq
Solving for $h$ one has
\beq
h^\pm(\sigma,r_d) \equiv 
 -\cos\sigma \pm \sqrt{\cos^2\sigma+r_d^4-1} \,.
\eeq
It follows that $h \in [h^{min}_{r_d},h^{max}_{r_d}]$ 
where 
\beq \label{hminmax}
h^{min}_{r_d}(\sigma) \equiv \mbox{max}(0,h^-(\sigma,r_d^{max}))
\andthis
h^{max}_{r_d}(\sigma) \equiv h^+(\sigma,r_d^{max})
\eeq
and that $h$ is excluded from the interval $[h^-_{r_d},~h^+_{r_d}]$,
where
\beq
h^-_{r_d}(\sigma) \equiv \mbox{max}(0,h^-(\sigma,r_d^{min})) 
\andthis
h^+_{r_d}(\sigma) \equiv h^+(\sigma,r_d^{min}) \,.
\eeq
The function $h^{max}_{r_d}(\sigma)$ corresponds to the upper solid
curves in Fig.~\ref{h-sigma-a} and Fig.~\ref{h-sigma-b}, while the
functions $h^\pm_{r_d}(\sigma)$ combine to the closed dotted curve
that excludes the area around $h=1$ and $\sigma=\pi$.  Due to the
stronger bound on $r_d^{min}$ in scenario (b) the corresponding
excluded area is larger as can be seen in Fig.~\ref{h-sigma-b}. In
scenario (c) there are two possible intervals for $r_d$, one
corresponding to the large band between the solid curves for positive
$\rho$ and one corresponding to the area between the dotted curves for
negative $\rho$ (see Fig.~\ref{h-sigma-c}).

\subsubsection{The $\theta_d$ constraint}
To make use of the upper and lower bounds on $\theta_d$ in
\eqs{rd-theta-a}--(\ref{rd-theta-c}) we eliminate $r_d$ from the
\eq{h-sigma}, getting:
\beq \label{sincosth}
\sin2\theta_d={h\sin\sigma\over \sqrt{h^2 + 2h \, \cos\sigma+1}} \andthis 
\cos2\theta_d={1+h \, \cos\sigma\over \sqrt{h^2 + 2h\, \cos\sigma+1}} \,.
\eeq
Solving for $h$ yields
\beq \label{h-sigma-theta2}
h^\pm(\sigma,\theta_d) = 
{\sin2\theta_d \, \sin(2\theta_d \pm \sigma)
 \over \sin^2\sigma -\sin^2 2\theta_d} \,.
\eeq
In order to find the upper and lower bound on $h$ as a function of
$\sigma$ from the global extrema of $\theta_d$ in
\eqs{rd-theta-a}--(\ref{rd-theta-c}) we have to treat carefully the
discrete ambiguities of the trigonometric functions in
\eq{h-sigma-theta2}. To be specific in the following we discuss the
constraints in scenario (a) and (b) that arise from
\eq{beta-bound}. For scenario (c) one has to consider separately the
two allowed interval for $\theta_d$ due to \eq{beta-bound-c}.  We
investigate separately the following three cases:
\begin{enumerate}
\item $2\theta_d \in [-90^\circ,~0^\circ]$: From $2\theta_d >
  -32^\circ$ it follows that $\sin2\theta_d$ increases monotonically
  in the interval $[-0.56,~0]$. Due to \eq{sincosth} the negative sign
  of $\sin2\theta_d$ implies that also $\sin\sigma$ is negative and
  therefore $\sigma \in [180^\circ,~360^\circ]$.  Taking into account
  that $\cos2\theta_d \ge 0$ (and therefore $h \le -1/\cos \sigma$ if
  $\sigma \in [180^\circ,~270^\circ]$) we find that there is only an
  upper bound on $h$ as a function of $\sigma$:
  \beq \label {h-upper} 
  h \le h^{max}_{\theta_d}(\sigma) \equiv h^+(\sigma,\theta_d^{min}) 
  \mbox{~~for~~} \sigma \in [180^\circ,~360^\circ-|2\theta_d^{min}|] \,.
  \eeq
\item $2\theta_d \in [0^\circ,~180^\circ]$: In this regime there are
  no constraints on $\theta_d$. Since
  $\mbox{sign}(\sin2\theta_d)=\mbox{sign}(\sin\sigma)$ it follows that
  also for $\sigma \in [0^\circ,~180^\circ]$ we cannot derive a limit
  on $h$.
\item $2\theta_d \in [180^\circ,~270^\circ]$: From $2\theta_d <
  212^\circ$ it follows that $\sin2\theta_d$ increases monotonically
  in the interval $[-0.53,~0]$. Due to \eq{sincosth} the negative sign
  of $\sin2\theta_d$ implies that also $\sin\sigma$ is negative and
  therefore $\sigma \in [180^\circ,~360^\circ]$.  Taking into account
  that $\cos2\theta_d \le 0$ (and therefore $h \ge -1/\cos \sigma$ if
  $\sigma \in [270^\circ,~360^\circ]$) we find that there is only a
  lower bound on $h$:
  \beq \label {h-lower}
  h \ge h^{min}_{\theta_d}(\sigma) \equiv h^+(\sigma,\theta_d^{max}) 
  \mbox{~~for~~} \sigma \in [180^\circ,2\theta_d^{max}] \,.
  \eeq
\end{enumerate} 

Together the constraints in \eq{h-upper} and \eq{h-lower} exclude the
region between the dashed [$h^{min}_{\theta_d}(\sigma)$] and the
dashed-dotted [$h^{max}_{\theta_d}(\sigma)$] curves in
Fig.~\ref{h-sigma-a} and Fig.~\ref{h-sigma-b}. For scenario (c) the
analysis is similar, resulting in the dashed ($\rho>0$) and
dashed-dotted curves ($\rho<0$) in Fig.~\ref{h-sigma-c}.

{}


\newpage
\vspace{2.5cm}

\begin{center}
Tab.~1: Input values\\
\vspace{1.5cm}
\begin{tabular}{|c|c|c|}  
\hline
~Parameter~ & Value &ref.\cr 
\hline \hline
$V_{ud}$            & $0.9735$ &      \cite{CKM2000}  \cr
\hline
$V_{ub}$            & $0.00355 \pm 0.00036$    &    \cite{CKM2000}   \cr
\hline
$V_{cb}$            & $0.040 \pm 0.0016$    &  \cite{CKM2000}   \cr
\hline
$\Delta m_d$        & $(0.487 \pm 0.014$) ps$^{-1}$&\cite{CKM2000}  \cr
\hline
$\Delta m_{B_s}$    & $> 15$ ps$^{-1}$     & \cite{CKM2000}   \cr
\hline
$|\epsK |$          & $0.00228$ &\cite{CKM2000}  \cr
\hline
$\BR(K^+ \to \pi^+ \nu \bar\nu)$ & $(1.5^{+3.4}_{-1.2}) \times
10^{-10}$&\cite{Adler}\cr
\hline
$\BR(B\to X_s\nu\bar\nu)$ &$<6.4\cdot10^{-4}$  & \cite{bsnunu}  \cr
\hline
$\eta_1$            &  $1.38  \pm 0.2$         & \cite{CKM2000} \cr  
\hline
$\eta_2$            &  $0.57$   &  \cite{CKM2000} \cr  
\hline
$\eta_3$            &  $0.47$   & \cite{CKM2000}  \cr  
\hline
$\eta_B$            & $0.55$    & \cite{Buras2001}  \cr
\hline
$\eta_X$            & $0.994$   &\cite{Buras2001}   \cr
\hline
$\xi$               & $1.14 \pm 0.05$       &\cite{CKM2000}   \cr
\hline
$P_c$               & $0.31 \pm 0.05$       &\cite{Buras2001} \cr
\hline
$P_0(X)$            & $0.42 \pm 0.06$       & \cite{Buras2001} \cr
\hline
$\bar m_t(m_t)$     & $(166 \pm 5)$ GeV     & \cite{Buras2001}  \cr
\hline
$A$                 & $0.819 \pm 0.040$     & \cite{CKM2000}  \cr
\hline
$\lambda$           & $0.2237\pm0.0033$     & \cite{CKM2000} \cr
\hline  
$B_K$               & $0.87\pm0.13$ GeV     & \cite{CKM2000}    \cr
\hline  
$f_{B_d}\sqrt{B_{B_d}}$  & $0.23\pm0.025$ GeV  &  \cite{CKM2000}  \cr
\hline  
$\kappa_+$               &$4.11 \times 10^{-11}$& \cite{Buras2001} \cr 
\hline\hline
\end{tabular}
\end{center}

\newpage
\begin{center} 
Tab.~2: Input values for the SBLR model\\
\vspace{1.5cm}
\begin{tabular}{|c|c|c|} 
\hline
~Parameter~ & Value & ref. \cr
\hline \hline
$\bar m_u(M_Z)$         & $0.0023 \pm 0.00042$ GeV & \cite{mass}        \cr
\hline
$\bar m_d(M_Z)$         & $0.0047 \pm 0.0006$ GeV      & \cite{mass}     \cr
\hline
$\bar m_c(M_Z)$         & $0.677 \pm 0.056$ GeV     & \cite{mass}      \cr
\hline
$\bar m_s(M_Z)$         & $0.0934 \pm 0.012$ GeV  & \cite{mass}         \cr
\hline
$\bar m_t(M_Z)$         & $181 \pm 13$ GeV    & \cite{mass}       \cr
\hline
$\bar m_b(M_Z)$         & $3 \pm 0.11$ GeV      & \cite{mass}     \cr
\hline
$M_{W_2}$              & $8000 \pm 5000$ GeV    & \cite{Ball1}       \cr
\hline
$M_{H}$              &  $12000  \pm 2000$ GeV    & \cite{Ball1}       \cr
\hline
$\delta_K$           &  $0.35 \pm 0.1$      & \cite{Bar1}       \cr  
\hline
$B_B^S (m_b)/B_B (m_b)$           &  $1.2  \pm 0.2$    & \cite{Ball1}      \cr  
\hline
$t_1$           &  $0.231$       &  -  \cr  
\hline
$t_2$           &  $0.041\pm0.018$   &  -      \cr  
\hline
$t_3$           &  $0.016  \pm 0.0016$   &  -      \cr  
\hline
$\eta_4$           &  $5  \pm 1.5$     & \cite{Bar1}       \cr  
\hline\hline
\end{tabular}
\end{center}


\putFig{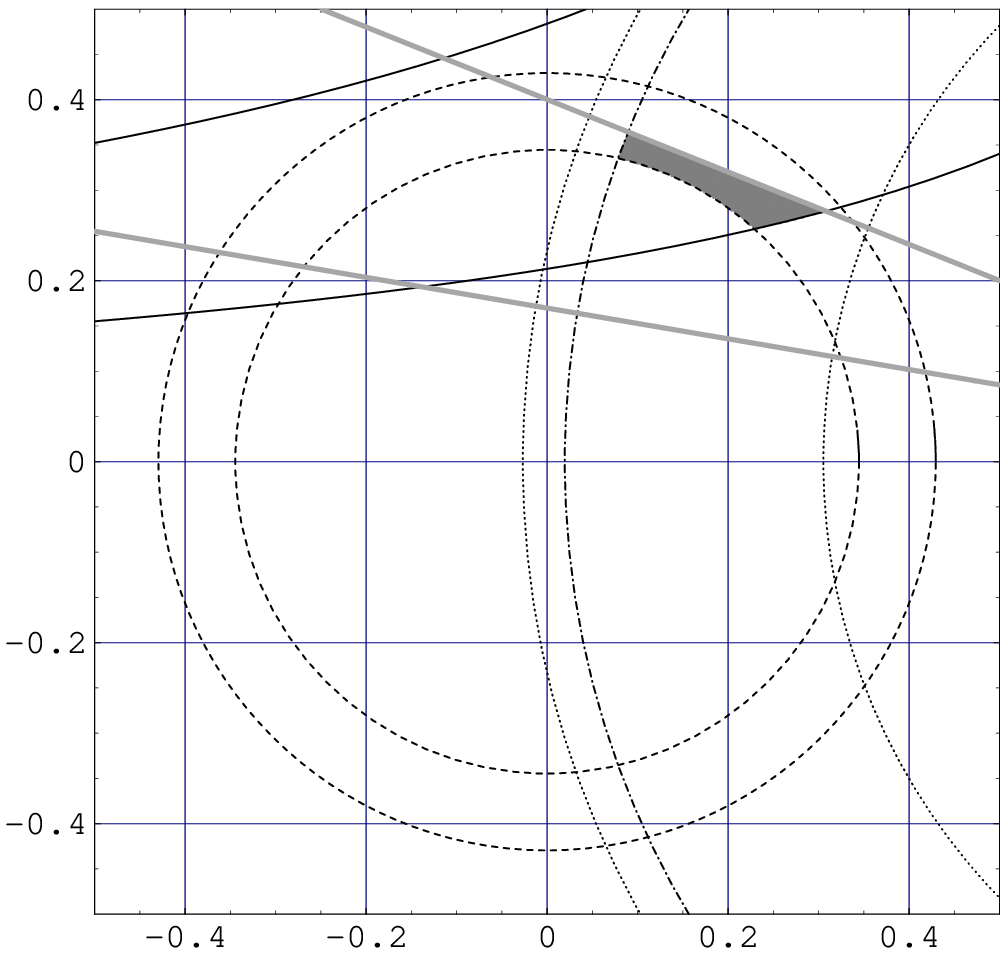}{\small Determination of the apex $(\rho,\eta)$ of
  the unitarity triangle: The grey area is the presently allowed
  region, which is determined by (a) the measurement of
  $|V_{ub}/V_{cb}|$ (dashed circles), (b) the observed $B_d^0 - \bar
  B_d^0$ mixing parameter $\Delta m_d$ (dotted circles), (c) the lower
  bound on the $B_s^0 - \bar B_s^0$ mixing parameter $\Delta m_s$
  (dashed-dotted circle), (d) the measurement of $\epsK$ (solid
  hyperbolae) and (e) the combined result of the CDF, BELLE and BaBar
  measurements of $a_{\psi K_S}$ (thick grey lines).}
{14}{14}{8}{\rho}{\eta}

\newpage  
\putFig{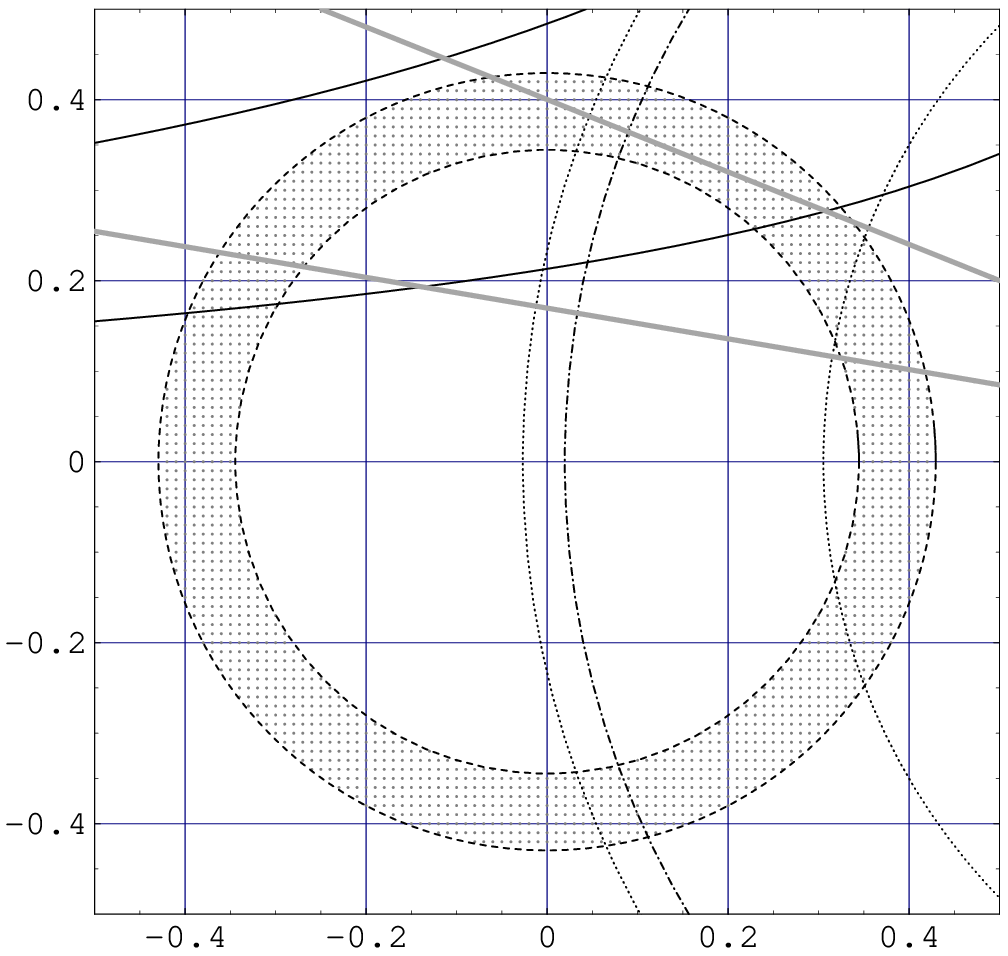}{The shaded area corresponds to the allowed region
  in the $\rho-\eta$ plane for scenario (a). It is constrained solely
  by $R_u \in [0.34,~0.43]$, which is determined by the measurement of
  $|V_{ub}/V_{cb}|$ (dashed circles). For comparison we also show the
  other constraints appearing in Fig.~\ref{rho-eta-SM}, which are not
  relevant here.}  {\widthAR}{\heightAR}{\spaAR}{\rho}{\eta}

\newpage \putFig{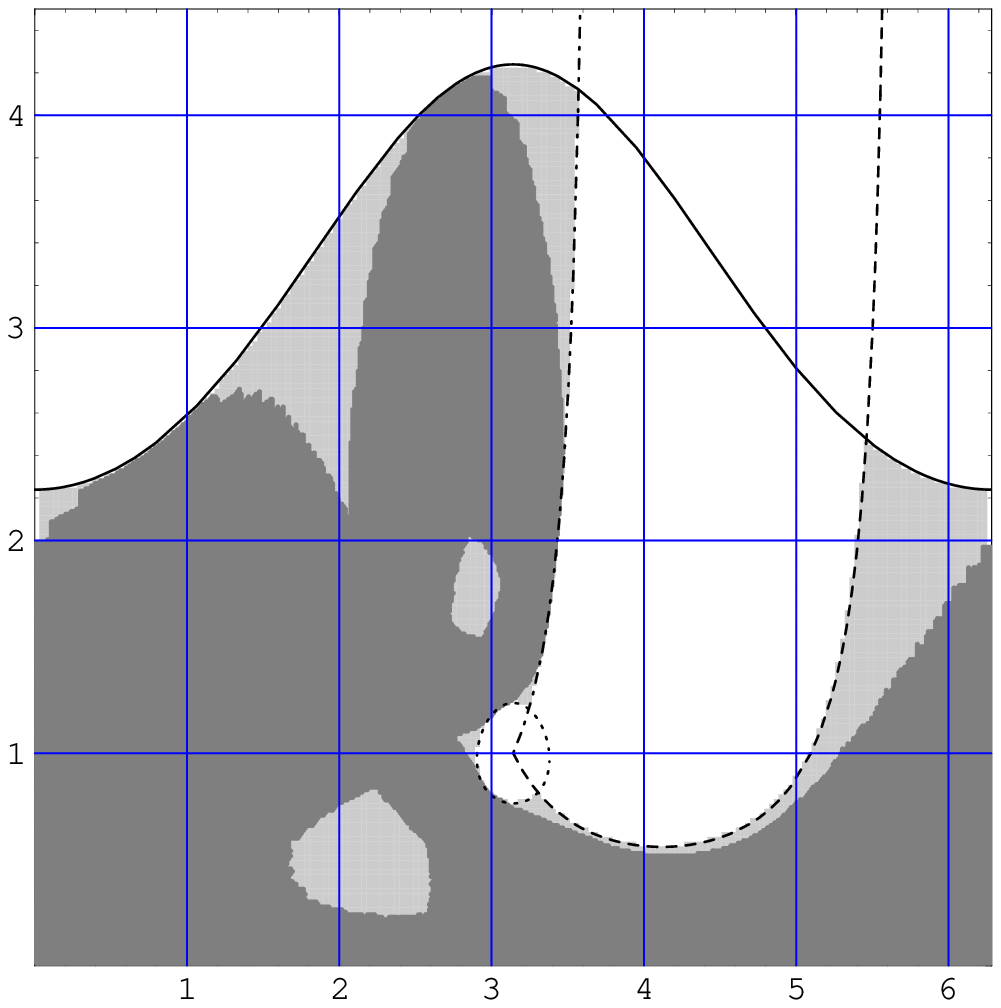}{The dark area corresponds to the allowed
  region in the $\h-\sigma$ plane for scenario (a) according to our
  numerical analysis. It is embedded in the light area that results
  from the analytical boundaries based on the global constraints on
  $r_d$ from $\Delta m_{d,s}$ and $\theta_d$ from $\apks$. Specifically
  the allowed range of $r_d$ excludes $h$ to be above
  $h^{max}_{r_d}(\sigma)$ (solid curve) or between
  $h^\pm_{r_d}(\sigma)$ (dotted curves), while the allowed range of
  $\theta_d$ exclude the region between $h^{min}_{\theta_d}(\sigma)$
  (dashed curve) and $h^{max}_{\theta_d}(\sigma)$ (dashed-dotted
  curve). The numerical results are not much more constraining than
  the analytical ones, except for the ``holes'' in the dark regions
  that correspond to the excluded region within the $R_u$ annulus,
  which has been ignored in ``global'' analysis. (See
  appendix~\ref{h-sig-details} for details.)}
{\widthAR}{\heightAR}{\spaAR}{\sigma}{h}

\newpage  
\putFig{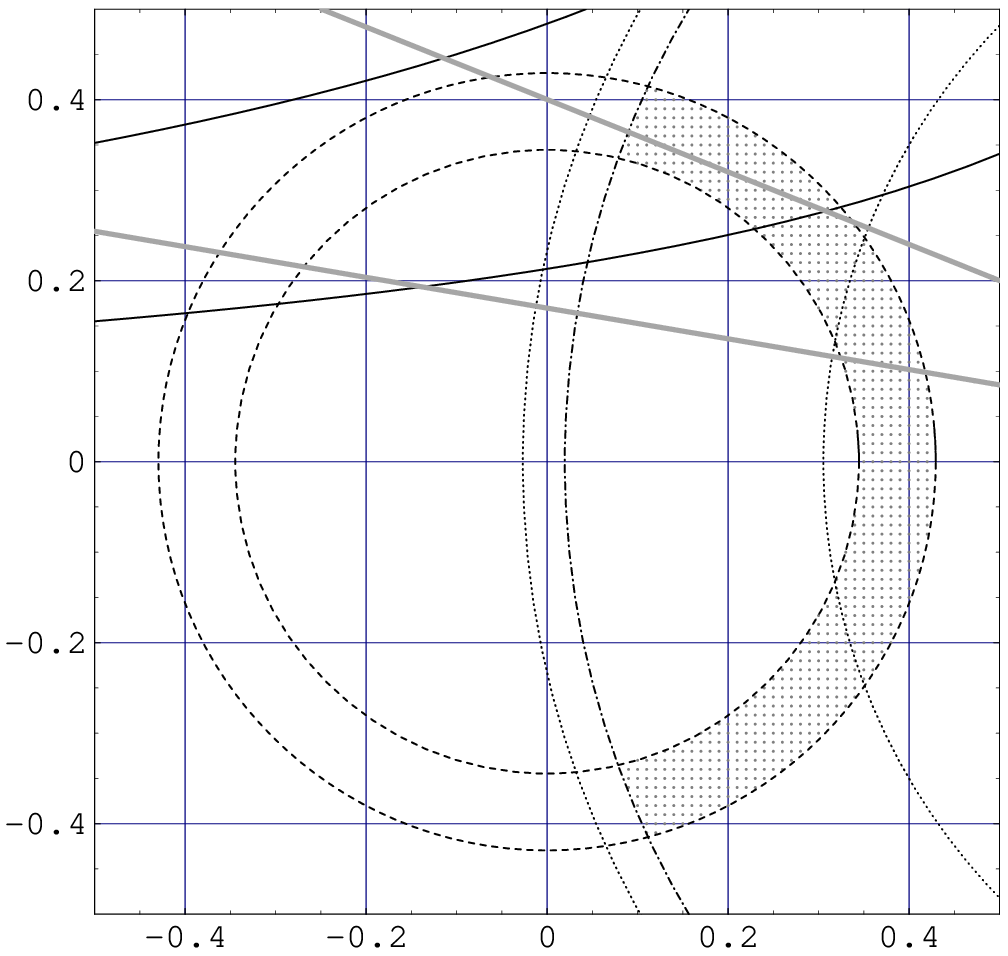}{The shaded area corresponds to the allowed region
  in the $\rho-\eta$ plane for scenario (b). It is constrained by $R_u
  \in [0.34,~0.43]$ (dashed circles) and $\Delta m_s/\Delta m_b$
  (dashed-dotted circle).  For comparison we also show the other
  constraints appearing in Fig.~\ref{rho-eta-SM}, which are not
  relevant here.}  {\widthAR}{\heightAR}{\spaAR}{\rho}{\eta}

\newpage \putFig{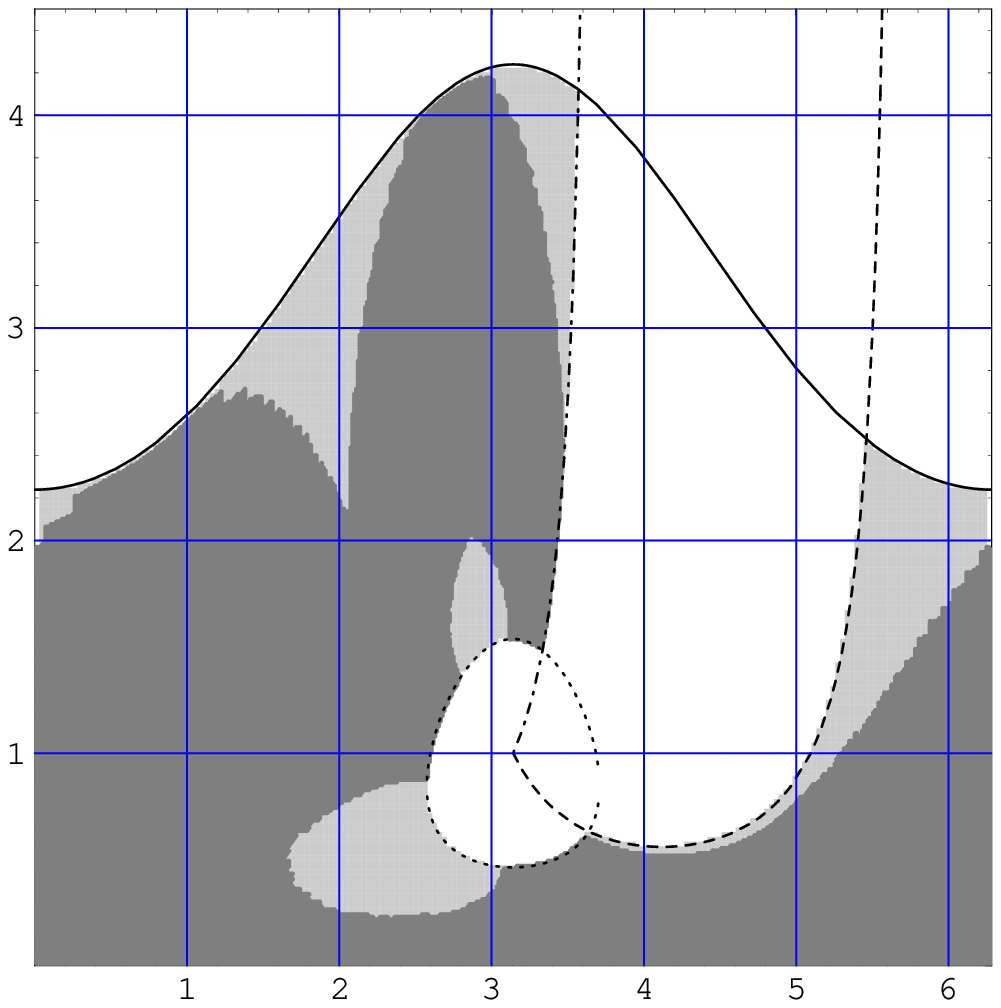}{The dark area corresponds to the allowed
  region in the $\h-\sigma$ plane for scenario (b) according to our
  numerical analysis. It is embedded in the light area that results
  from the analytical boundaries based on the global constraints on
  $r_d$ from $\Delta m_{d,s}$ and $\theta_d$ from $\apks$. Specifically
  the allowed range of $r_d$ excludes $h$ to be above
  $h^{max}_{r_d}(\sigma)$ (solid curve) or between
  $h^\pm_{r_d}(\sigma)$ (dotted curves), while the allowed range of
  $\theta_d$ excludes the region between $h^{min}_{\theta_d}(\sigma)$
  (dashed curve) and $h^{max}_{\theta_d}(\sigma)$ (dashed-dotted
  curve). Due to the stronger bound on $r_d^{min}$ in scenario (b) the
  excluded area from $h^\pm_{r_d}(\sigma)$ is larger than in
  Fig.~\ref{h-sigma-a}. (See appendix~\ref{h-sig-details} for
  details.)} {\widthAR}{\heightAR}{\spaAR}{\sigma}{h}

\newpage  
\putFig{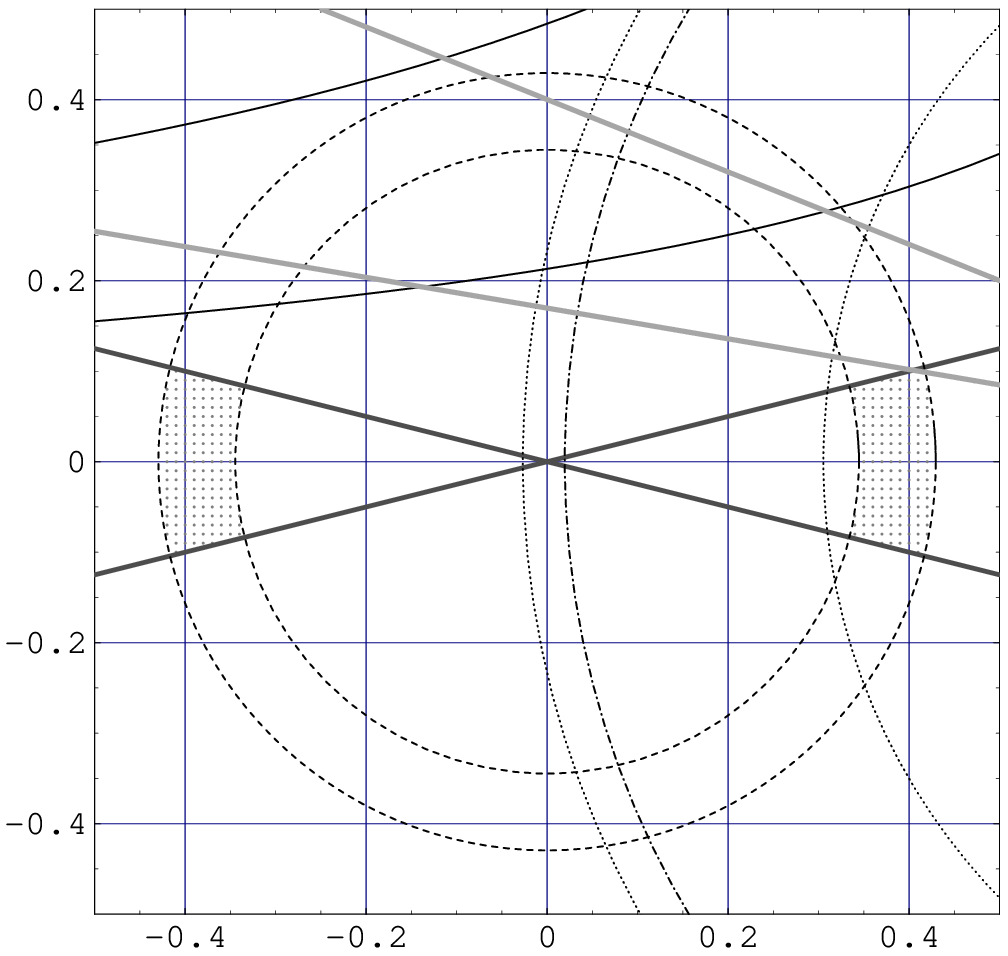}{The shaded area corresponds to the allowed region
  in the $\rho-\eta$ plane for scenario (c). It is constrained by $R_u
  \in [0.34,~0.43]$ (dashed circles) and $\tan \delta_{KM} < 0.25$
  (``small phase'').  For comparison we also show the other
  constraints appearing in Fig.~\ref{rho-eta-SM}, which are not
  relevant here.}  {\widthAR}{\heightAR}{\spaAR}{\rho}{\eta}

\newpage \putFig{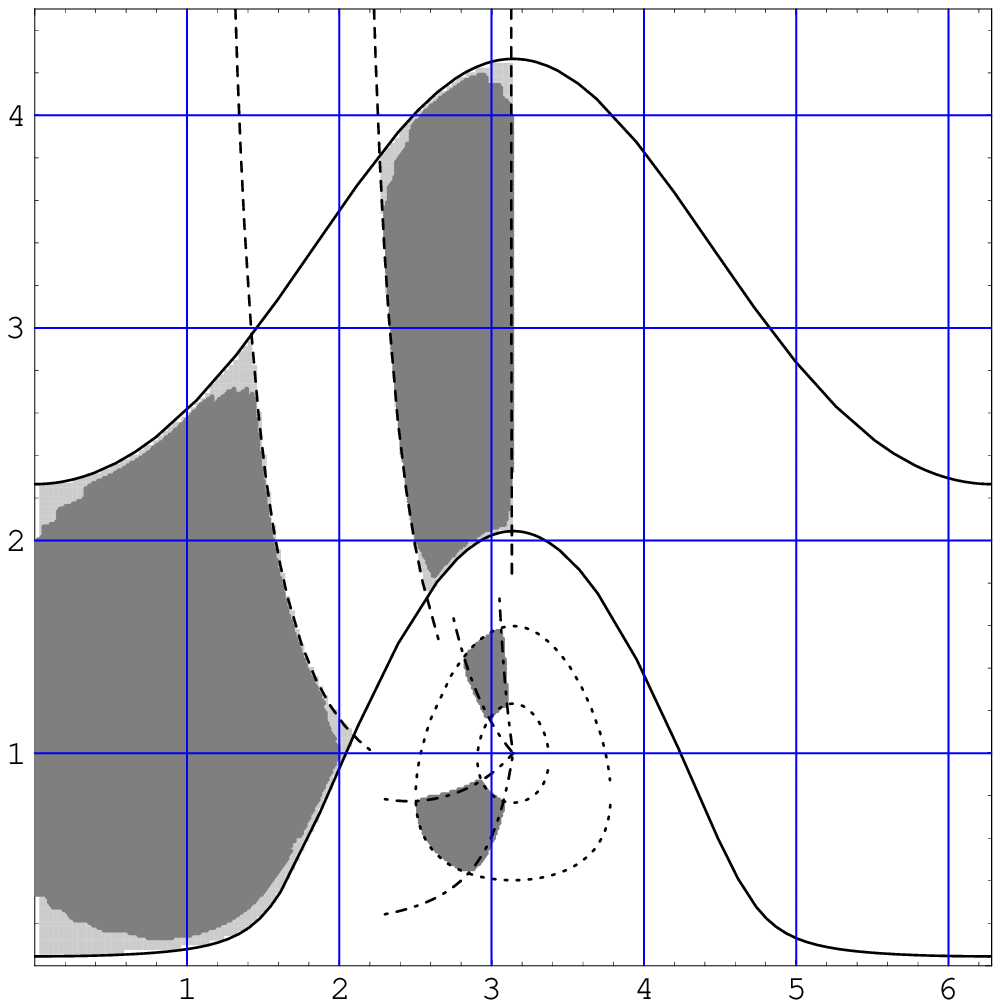}{The dark area corresponds to the allowed
  region in the $\h-\sigma$ plane for scenario (c) according to our
  numerical analysis. It is embedded in the light area that results
  from the analytical boundaries based on the global constraints on
  $r_d$ from $\Delta m_{d,s}$ and $\theta_d$ from $\apks$. Specifically
  there are two possible intervals for $r_d$, one corresponding to the
  large band between the solid curves for positive $\rho$ and one
  corresponding to the area between the dotted curves for negative
  $\rho$. The allowed ranges of $\theta_d$ limit the allowed region to
  be between the dashed ($\rho>0$) and dashed-dotted curves
  ($\rho<0$).  Note that unlike for scenarios (a) and (b) in scenario
  (c) $h$ has a lower bound and $\sigma$ has an upper bound. (See
  appendix~\ref{h-sig-details} for details.)}
{\widthAR}{\heightAR}{\spaAR}{\sigma}{h}
  
\newpage 

\putFig{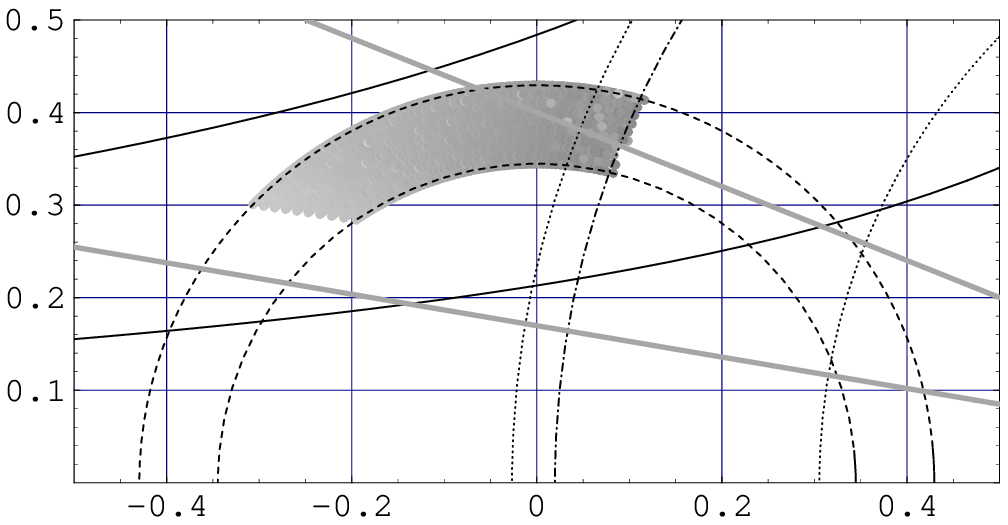}{The shaded area corresponds to the ``first'' solution
  resulting from \eqs{Rb}, (\ref{MFV-DelMb}) and~(\ref{MFV-eps})
  within models of MFV. The grey scale indicates the magnitude of
  $F_{tt}$ as a function of $\rho$ and $\eta$: The lightest region
  corresponds to $F_{tt} \simeq 0.4$ and the darkest region
  corresponds to $F_{tt} \simeq 2.7$. Adding the constraint from
  $\Delta m_s/\Delta m_b$ (dashed-dotted circle) practically the
  entire solution presented in this figure is excluded.}
  {11}{5.5}{\spaAR}{\rho}{\eta}

\putFig{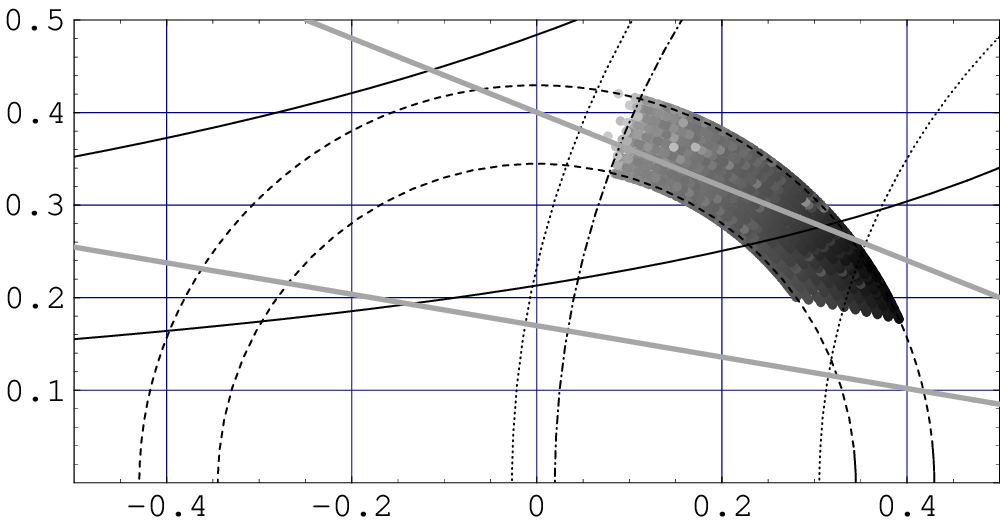}{The shaded area corresponds to the allowed region
  according to the ``second'' solution resulting from \eqs{Rb},
  (\ref{MFV-DelMb}) and~(\ref{MFV-eps}) within models of MFV. The grey
  scale indicates the magnitude of $F_{tt}$ as a function of $\rho$
  and $\eta$: The lightest region corresponds to $F_{tt} \simeq 1.2$
  and the darkest region corresponds to $F_{tt} \simeq 5.7$.}
  {11}{5.5}{\spaAR}{\rho}{\eta}


\end{document}